\documentclass[twocolumn]{aastex631}

\usepackage{graphicx}	
\usepackage{amsmath}	
\usepackage{subfigure}
\usepackage{natbib}
\usepackage{color}
\usepackage{tabularx}
\usepackage{longtable}
\usepackage{bm}
\usepackage{threeparttable}
\usepackage{enumerate}
\usepackage[normalem]{ulem}
\usepackage{verbatim}
\usepackage[T1]{fontenc}

\newcommand{\code}[1]{{\texttt{#1}}}
\newcommand{\tess}{{\it TESS}}

\newcommand{\tar}{{TOI-4201}}

\begin{document}

\title{The Aligned Orbit of a Hot Jupiter around the M Dwarf TOI-4201}

\shorttitle{The Spin-Orbit Alignment of the TOI-4201 system}
\shortauthors{Gan et al.}

\correspondingauthor{Tianjun Gan}
\email{tianjungan@gmail.com}

\author[0000-0002-4503-9705]{Tianjun~Gan}
\affil{Department of Astronomy, Tsinghua University, Beijing 100084, People's Republic of China}

\author[0000-0002-6937-9034]{Sharon X. Wang}
\affil{Department of Astronomy, Tsinghua University, Beijing 100084, People's Republic of China}

\author[0000-0002-8958-0683]{Fei Dai}
\affil{Institute for Astronomy, University of Hawaii at Manoa, Honolulu, HI 96822, USA}

\author[0000-0002-4265-047X]{Joshua~N.~Winn}
\affil{Department of Astrophysical Sciences, Princeton University, 4 Ivy Lane, Princeton, NJ 08544, USA}

\author[0000-0001-8317-2788]{Shude Mao}
\affil{Department of Astronomy, Tsinghua University, Beijing 100084, People's Republic of China}

\author[0000-0002-8808-4282]{Siyi Xu}
\affil{Gemini Observatory/NSF NOIRLab, 670 N. A'ohoku Place, Hilo, HI 96720, USA}

\author[0000-0003-0987-1593]{Enric Pall{\'e}}
\affil{Instituto de Astrof\'isica de Canarias (IAC), V\'ia L\'actea s/n, E-38205 La Laguna, Tenerife, Spain}
\affil{Dept. Astrof\'sica, Universidad de La Laguna (ULL), E-38206 La Laguna, Tenerife, Spain}

\author[0000-0003-4733-6532]{Jacob L. Bean} 
\affil{Department of Astronomy \& Astrophysics, University of Chicago, 5640 South Ellis Avenue, Chicago, IL 60637, USA}

\author[0000-0003-2404-2427]{Madison Brady} 
\affil{Department of Astronomy \& Astrophysics, University of Chicago, 5640 South Ellis Avenue, Chicago, IL 60637, USA}

\author[0009-0003-1142-292X]{Nina Brown} 
\affil{Department of Astronomy \& Astrophysics, University of Chicago, 5640 South Ellis Avenue, Chicago, IL 60637, USA}

\author[0000-0001-9352-0248]{Cicero Lu} 
\affil{Gemini Observatory/NSF NOIRLab, 670 N. A'ohoku Place, Hilo, HI 96720, USA}

\author[0000-0002-4671-2957]{Rafael Luque} 
\affil{Department of Astronomy \& Astrophysics, University of Chicago, 5640 South Ellis Avenue, Chicago, IL 60637, USA}

\author[0000-0003-4603-556X]{Teo Mocnik} 
\affil{Gemini Observatory/NSF NOIRLab, 670 N. A'ohoku Place, Hilo, HI 96720, USA}

\author[0000-0003-4526-3747]{Andreas Seifahrt} 
\affil{Gemini Observatory/NSF NOIRLab, 670 N. A'ohoku Place, Hilo, HI 96720, USA}

\author[0000-0001-7409-5688]{Gu{\dh}mundur K. Stefánsson} 
\affil{Anton Pannekoek Institute of Astronomy, Science Park 904, University of Amsterdam, 1098 XH Amsterdam, Netherlands}



\begin{abstract}

Measuring the obliquities of stars hosting giant planets may shed light on the dynamical history of planetary systems. Significant efforts have been made to measure the obliquities of FGK stars with hot Jupiters, mainly based on observations of the Rossiter-McLaughlin effect. In contrast, M dwarfs with hot Jupiters have hardly been explored, because such systems are rare and often not favorable for such precise observations. Here, we report the first detection of the Rossiter-McLaughlin effect for an M dwarf with a hot Jupiter, TOI-4201, using the Gemini-North/MAROON-X spectrograph. We find TOI-4201 to be well-aligned with its giant planet, with a sky-projected obliquity of $\lambda=-3.0_{-3.2}^{+3.7}\ ^{\circ}$ and a true obliquity of $\psi=21.3_{-12.8}^{+12.5}\ ^{\circ}$ with an upper limit of $40^{\circ}$ at a 95\% confidence level. The result agrees with dynamically quiet formation or tidal obliquity damping that realigned the system. As the first hot Jupiter around an M dwarf with its obliquity measured, TOI-4201b joins the group of aligned giant planets around cool stars ($T_{\rm eff}<6{,}250$~K), as well as the small but growing sample of planets with relatively high planet-to-star mass ratio ($M_p/M_\ast\gtrsim 3\times 10^{-3}$) that also appear to be mostly aligned. 

\end{abstract}



\section{Introduction} \label{sec:intro}

Since the first discovery of a hot Jupiter \citep{Mayor1995}, the origin of such short-period planets has been the subject of much research.
Three basic hypotheses have been postulated: {\it in-situ} formation, disk-driven migration, and high-eccentricity migration
\citep{Dawson2018}. Unlike the other two scenarios, high-eccentricity migration would tend to excite inclinations
through planet scattering \citep{Rasio1996,Ford2008,Chatterjee2008}, Kozai-Lidov interactions \citep{Fabrycky2007,Naoz2016} and secular resonances  \citep{Wu2011,Petrovich2020}. All these routes might be expected to result in misalignment between the planet's orbital angular momentum vector and the star's spin angular momentum vector (i.e., a large stellar obliquity). For this reason, measuring stellar obliquities is useful
as a probe of the dynamical history of close-orbiting
giant planets \citep[see][and references therein]{Albrecht2022}. 

One way to determine the stellar obliquity is through the Rossiter-McLaughlin (RM) effect \citep{Rossiter1924,McLaughlin1924}, the distortion in the stellar spectral lines due to selective blockage of the
star's rotating photosphere by a transiting planet. Unlike the Sun, for
which the equatorial plane is tilted by only $7^{\circ}$ from the ecliptic \citep{Beck2005}, a significant fraction of hot-Jupiter hosts are found to be misaligned \citep{Winn2015}. In particular, relatively hot and massive stars ($T_{\rm} > 6{,}250$~K, $M> 1.3\,M_\odot$) exhibit a broad range of spin–orbit angles while cooler and less massive stars tend to be well-aligned \citep{Schlaufman2010,Winn2010}. The critical effective temperature separating these groups is close to the ``Kraft break'' \citep{Kraft1967} that separates stars with convective and radiative envelopes. Since cool stars have thicker and more massive convective envelopes than hot stars \citep{Pinsonneault2001}, and convective envelopes
are thought to allow for more rapid tidal dissipation,
\cite{Winn2010} speculated that many hot Jupiters once had
misaligned orbits but tidal dissipation damped the obliquities and realigned the systems \citep[see also][]{Albrecht2012,Wang2021K2_232,Spalding2022}. If so, then one might expect hot Jupiters around M dwarfs to be especially well aligned, since they have deep convective zones. One might also wonder whether late M dwarfs, which are fully convective, should allow for rapid tidal obliquity damping or if the absence of a radiative/convective boundary changes the situation.

There have been many recent studies of orbital misalignment of M dwarfs
with non-giant planets, such as TRAPPIST-1 \citep{Hirano2020a,Brady2023},
GJ\,436 \citep{Bourrier2018}, AU\,Mic
\citep{Hirano2020b,Palle2020,Martioli2020,Addison2021}, K2-25 \citep{Stefansson2020}, GJ\,3470 \citep{Stefansson2022} and K2-33 \citep{Hirano2024}. However, obliquity measurements for M dwarfs with hot Jupiters are lacking due mainly to the low occurrence rate of such systems \citep{Gan2023,Bryant2023}. To our knowledge, \cite{Dai2018kepler45} made the first attempt to measure the obliquity of an M dwarf with a hot Jupiter, Kepler-45b. Instead of using the
RM effect, they studied the light-curve anomalies produced
when a transiting planet crosses over starspots \citep{Sanchis-Ojeda2011wasp4}, and found an upper limit of $10^{\circ}$ on the obliquity of the host star. With a similar methodology, recent works by \cite{Almenara2022} and \cite{LibbyRoberts2023} both found that TOI-3884, an M4 dwarf hosting a super-Neptune, is likely misaligned.

Further progress has been made possible by the full-sky photometric survey performed by NASA's Transiting Exoplanet Survey Satellite \citep[TESS;][]{Ricker2015}, which has led to the enlargement of the sample of M dwarfs
with hot Jupiters. Alongside this development is the advent of a new generation of high-resolution stabilized spectrographs on large telescopes.
Together, these advances provide an opportunity to extend obliquity studies to M dwarfs with giant planets. Here, we present the first measurement of the RM effect for a hot Jupiter transiting an M dwarf, TOI-4201b. The star is an early M dwarf at a distance about 189 pc, and it hosts a hot Jupiter with an orbital period of 3.58 days \citep{Gan2023TOI4201,Hartman2023,Delamer2024}. The rest of this Letter is organized as follows: Section~\ref{maroonx_data} details the spectroscopic observations, Section~\ref{jointfit} presents the joint-fit analysis, Section~\ref{discussion} discusses the results,
and Section~\ref{conclusion} describes our conclusions.

\section{MAROON-X spectroscopic observations}\label{maroonx_data}

Since the host star is faint ($V=15.3$ and $J=12.3$) and the transit duration is relatively short ($\sim 2$ hours), high-resolution spectroscopic instruments on large telescopes are required to achieve high RV precision in a short amount of time while having enough sampling. We collected 19 spectra with an exposure time of 900s on UT 2023 December 26 using MAROON-X under the program GN-2023B-FT-107, covering a full transit and a total of 3 hours outside of the transit. MAROON-X is a high-resolution ($R\sim85,000$) optical fiber-fed echelle spectrograph installed on the 8.1\,m Gemini North Telescope, on Maunakea, Hawaii, with a wavelength range from 500 to 920 nm \citep{Seifahrt2018,Seifahrt2020}. During the observations, the airmass varied between 1.2 and 1.8, the sky was clear, and the seeing was around $0.4\arcsec$ but degraded to $0.7\arcsec$ after egress. The  signal-to-noise ratio (SNR) per resolution element near the $H\alpha$ line (Order 93) is about 19 at the beginning, but decreasing to 14 after egress due to degraded seeing. 

The raw MAROON-X data were reduced with custom Python 3 routines based on routines originally written for the CRIRES instrument \citep{Bean2010}. We then utilized the SpEctrum Radial Velocity AnaLyser \citep[\code{SERVAL};][]{Zechmeister2018} pipeline to measure the radial velocities (RVs) based on the template matching method. We separately obtained RV estimates from the blue (500-670 nm) and red (650-920 nm) arms of the spectrograph, after correcting the main instrumental drift, and we treated these two RV time series as though they
were from two different instruments. We converted all the time stamps of our measurements from JD to BJD \citep{Eastman2010}. In addition to the RVs, we extracted stellar activity diagnostics including the chromatic RV index (CRX), the differential line width (dLW), and the H$\alpha$ activity indices \citep{Zechmeister2018}. Both the red and blue arms captured the $\rm H\alpha$ line, allowing for independent measurements.
Table~\ref{RVtable} gives all of the results.
The median uncertainties of the RVs from the blue and red arms are 4.5 and 4.6~$\rm m\ s^{-1}$, respectively.

To rule out the possibility that short-term stellar activity (i.e., flares) mimic the RM signal, we examined the correlation between the activity indices and the apparent RVs, quantified by the Pearson Correlation Coefficient. We found no evidence for significant correlations ($p<0.05$).

\begin{figure*}[ht]
    \centering
    \includegraphics[width=0.99\linewidth]{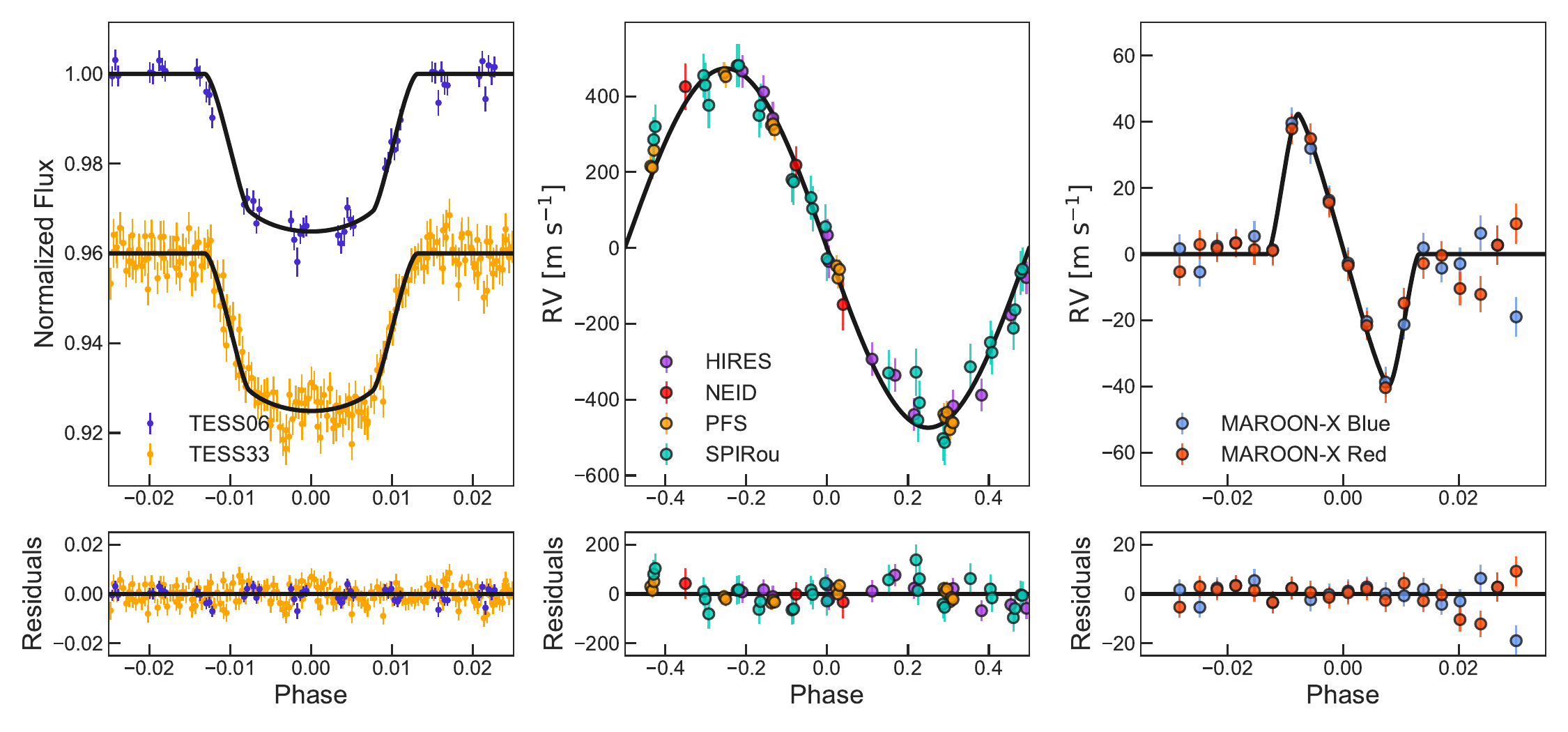}
    \caption{TESS transit photometry of \tar\ from Sectors 06 and 33 (left), out-of-transit RVs (middle), and RM measurements after subtracting the best-fit Keplerian model (right). The excess scatter in the RVs after egress is probably due to the degraded seeing. In each case, the black curve illustrates the best-fit model. The plotted error bars are the quadrature sums of the formal measurement uncertainties and the fitted ``jitter'' parameters. Residuals are shown in the bottom panels.}
    \label{fig:joint-fit}
\end{figure*}

\begin{figure*}[ht]
    \centering
    \includegraphics[width=0.99\linewidth]{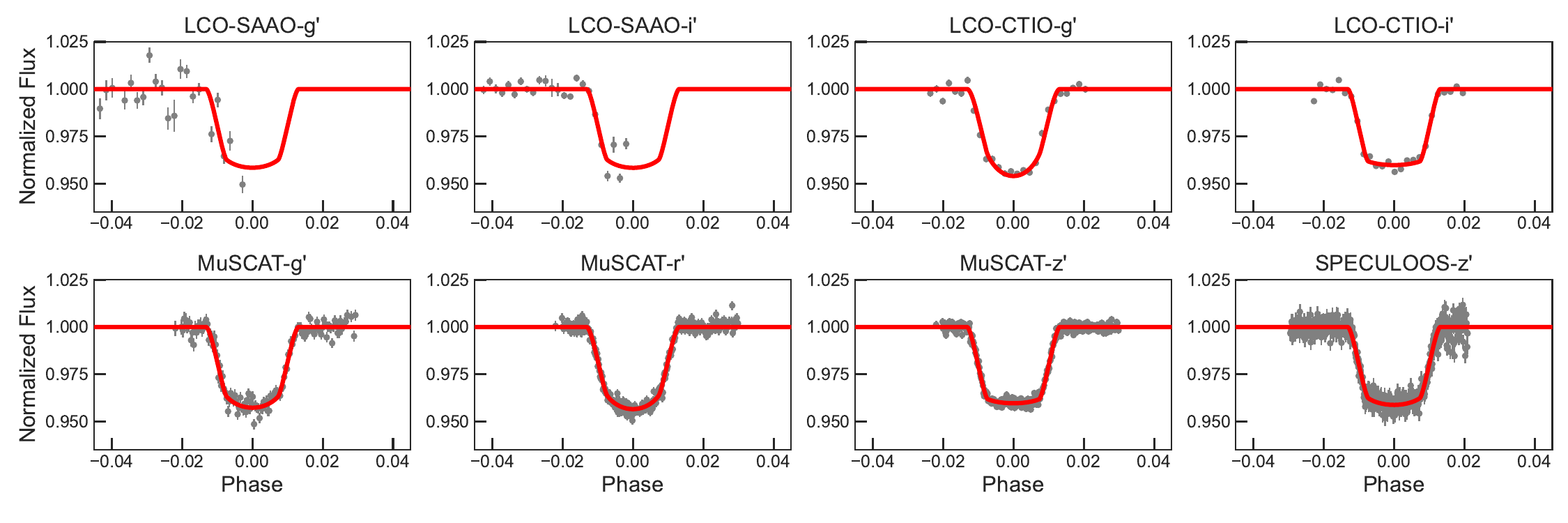}
    \caption{Dimensionless phase-folded ground-based transit observations of \tar\ with maximum phase of 1. The instrument name and observation filter are shown at the top of each panel. The red solid lines represent the best-fit transit models from the joint-fit analysis.}
    \label{fig:ground-fit}
\end{figure*}

\section{Analysis}\label{jointfit}

We employed the \code{Allesfitter} code \citep{Gunther2021} to measure the sky-projected spin-orbit angle ($\lambda$) for \tar b. We performed a joint fit of two \tess\ light curves from Sector 06 and 33 with cadences of 30 and 10 mins; eight ground-based light curves from LCOGT, MuSCAT and SPECULOOS publicly available on ExoFOP\footnote{\url{https://exofop.ipac.caltech.edu/tess/}}; RV data drawn from the literature; and our RM measurements. The TESS light curves are adopted from \cite{Gan2023TOI4201}, which were extracted through simple aperture photometry. The out-of-transit RVs were from CFHT/SPIRou \citep{Gan2023TOI4201},
Keck/HIRES \citep{Hartman2023},
WIYN/NEID, and Magellan/PFS (both from \citealt{Delamer2024}).

The joint model has ten key parameters: orbital period ($P$), mid-transit time ($T_{0}$), planet-to-star radius ratio ($R_p/R_\ast$), sum of radii divided by the orbital semi-major axis (($R_p+R_\ast)/a$), cosine of the orbital inclination ($\cos i_p$), RV semi-amplitude ($K$), eccentricity parameters ($\sqrt{e} \cos \omega$ and $\sqrt{e} \sin \omega$), sky-projected spin-orbit angle ($\lambda$) and projected rotational velocity ($v\sin i$). Moreover, we allowed for a light dilution factor\footnote{$\rm D=F_{C}/\left(F_{T}+F_{C}\right)$, where $\rm F_T$ and $\rm F_C$ represent the target and contamination fluxes.} for the TESS data due to the large pixel scale ($21\arcsec$/pixel).
This was not necessary for the ground-based observations, for which \tar\ was well resolved. We adopted a quadratic limb-darkening law for the TESS photometry and a linear law for the ground-based data \citep{Kipping2013}. The simpler linear law was adopted for the ground-based data because the amount of data from each instrument and the SNR was limited compared with TESS. For each photometric and spectroscopic dataset, we allowed for a baseline offset and a ``jitter'' term to account for unmodeled sources of white noise. Uniform priors were placed on all parameters except for the TESS dilution factor, for which we set a truncated normal prior. 

We performed an affine-invariant Markov Chain Monte Carlo (MCMC) analysis with 140 walkers to sample the posterior distributions of all parameters. A total of 150{,}000 steps were taken by each walker, and the first 30{,}000 ``burn-in'' steps were excluded. All Markov chains were run for more than 30 times their autocorrelation length so that the convergence was reached \citep{Foreman2013}. Figure~\ref{fig:joint-fit} and Table~\ref{allpriors} summarize the main results.
Among the results, we obtained $\lambda=-3.0_{-3.2}^{+3.7}\ ^{\circ}$, indicating a good alignment. The ground-based light curves along with the best-fit transit models are presented in Figure~\ref{fig:ground-fit}. The posteriors of other relevant parameters are shown in Table~\ref{allotherpriors} of the Appendix.

Based on the periodogram analysis for TESS data and ground-based long-term light curves from the Zwicky Transient Facility \citep[ZTF;][]{Bellm2019,Masci2019}, \cite{Gan2023TOI4201} reported that \tar\ shows a $17.3\pm0.4$ days photometric modulation, which is likely to be
from stellar rotation. Assuming this
is the case, and neglecting the effects
of differential rotation, the stellar equatorial rotation velocity
$v=2\pi R_{\ast}/P_{\rm rot} = 1.8\pm0.2$ km~s$^{-1}$, which is
consistent with the measured value
of $v\sin i = 1.65^{+0.11}_{-0.09}$~km/s within about $1\sigma$.
Thus, there is no strong evidence for the
difference between the stellar inclination and the orbital inclination angles.
Nevertheless, we used the Bayesian inference methodology proposed by \cite{Masuda2020} to place a quantitative constraint on $i_{\star}$,
based on the likelihood function\begin{equation}\label{tau}
\begin{aligned}
    \mathcal{L} = & \left(\frac{R_\ast/R_{\odot}-0.63}{0.02}\right)^{2}+\left(\frac{P_{\rm rot}-17.3\ {\rm days}}{0.4\ {\rm days}}\right)^{2} \\
    &+\left(\frac{v\sqrt{1-\cos^{2} i_\star}-1.6\ {\rm km/s}}{0.2\ {\rm km/s}}\right)^{2}.
\end{aligned}
\end{equation}
We set uniform priors on $R_{\ast}$ ($0\leq R_{\ast}\leq 10\ R_\odot$), $P_{\rm rot}$ ($0\leq P_{\rm rot}\leq 100$ days) and the cosine of stellar inclination $\cos i_{\star}$ ($0\leq \cos i_\star\leq 1$), and sampled the parameter space using \code{emcee} \citep{Foreman2013}. We initialized 150 walkers that each took 50{,}000 steps, and discarded the first 5{,}000 samples. We obtained a posterior of $\cos i_{\star}=0.38^{+0.19}_{-0.24}$, or a stellar inclination of $i_\star=67.4_{-12.8}^{+14.3}\ ^{\circ}$. The 3D obliquity ($\psi$) of \tar\ is then determined via \citep{Albrecht2022}
\begin{equation}
    \cos \psi = \cos i_{\star}\cos i_{p}+\sin i_{\star}\sin i_{p}\cos \lambda.
\end{equation}
We used the posteriors of $i_\star$, $i_p$ and $\lambda$ from the analysis above to obtain the true obliquity. The resulting true obliquity is $\psi=21.3_{-12.8}^{+12.5}\ ^{\circ}$ with a 95\% confidence that $\psi \leq 40^{\circ}$, indicating that the system is aligned.

\begin{table*}\scriptsize
    {\renewcommand{\arraystretch}{1.1}
    \caption{Parameter priors and best-fits in the joint model for \tar. $\mathcal{U}$(a, b) stands for a uniform prior between $a$ and $b$.}
    \begin{tabular}{lccr}
        \hline\hline
        Parameter       &Prior &Best-fit    &Description\\\hline
        \it{Stellar parameters$^{[1]}$}\\
        $M_\ast$ ($M_\odot$) &$\cdots$ &$0.61\pm0.02$ &Stellar mass\\
        $R_\ast$ ($R_\odot$) &$\cdots$ &$0.63\pm0.02$ &Stellar radius\\
        $T_{\rm eff}$ (K) &$\cdots$ &$3794\pm79$ &Stellar effective temperature\\
        $\log g_\ast$ (cgs) &$\cdots$ &$4.64\pm0.03$ &Stellar surface gravity\\
        $P_{\rm rot}$ (days) &$\cdots$ &$17.3\pm0.4$ &Stellar rotation period \\\hline
        \it{Key fitted parameters}\\
        $P$ (days)  &$\mathcal{U}$ ($3.0$\ ,\ $4.0$)  &$3.5819198_{-0.0000012}^{+0.0000013}$
        &Orbital period\\
        $T_{0}$ (BJD-2457000) &$\mathcal{U}$ ($1470.93$\ ,\ $1470.99$) &$1470.9618_{-0.0003}^{+0.0004}$ &Mid-Transit time\\
        $R_p/R_\ast$ &$\mathcal{U}$ ($0.0$\ ,\ $0.5$)  &$0.1949_{-0.0012}^{+0.0013}$
        &Planet-to-star radius ratio\\
        $(R_p+R_\ast)/a$ &$\mathcal{U}$ ($0.0$\ ,\ $0.5$)  &$0.0890_{-0.0013}^{+0.0015}$
        &Sum of radii divided by the orbital semi-major axis\\
        $\cos i_p$ &$\mathcal{U}$ ($0.0$\ ,\ $1.0$)  &$0.0361_{-0.0023}^{+0.0025}$
        &Cosine of the orbital inclination\\
        $\sqrt{e}\cos \omega$ &$\mathcal{U}$ ($-1$\ ,\ $1$) &$-0.045_{-0.066}^{+0.076}$ &Parametrization for $e$ and $\omega$\\
        $\sqrt{e}\sin \omega$ &$\mathcal{U}$ ($-1$\ ,\ $1$) &$-0.076_{-0.071}^{+0.083}$ &Parametrization for $e$ and $\omega$\\
        $K$ (m~s$^{-1}$) &$\mathcal{U}$ ($0$\ ,\ $1000$) &$478.5_{-7.1}^{+6.7}$ &RV semi-amplitude\\
        $v\sin i$ (km s$^{-1}$) &$\mathcal{U}$ ($0.1$\ ,\ $10$) &$1.65_{-0.09}^{+0.11}$ &Projected stellar rotation velocity\\
        $\lambda$ (deg) &$\mathcal{U}$ ($-180$\ ,\ $180$) &$-3.0_{-3.2}^{+3.7}$ &Projected spin-orbit angle\\\hline
        \it{Derived stellar parameters}\\
        $i_\star$ (deg) &$\cdots$ &$67.4_{-12.8}^{+14.3}$ &Stellar inclination\\
        $\psi$ (deg) &$\cdots$ &$21.3_{-12.8}^{+12.5}$ &True obliquity\\\hline
        \it{Derived planetary parameters}\\
        $R_{p}$ ($R_{J}$) &$\cdots$ &$1.19_{-0.05}^{+0.04}$ &Planet radius\\
        $M_{p}$ ($M_{J}$) &$\cdots$ &$2.59_{-0.11}^{+0.10}$ &Planet mass\\
        $a$ (AU) &$\cdots$ &$0.0398_{-0.0022}^{+0.0020}$ &Semi-major axis\\
        $i_p$ (deg) &$\cdots$ &$87.9_{-0.2}^{+0.2}$ &Orbital inclination\\
        $e$ &$\cdots$ &$0.008_{-0.007}^{+0.026}$ &Orbital eccentricity\\
        $T_{\rm eq}^{[2]}$ (K) &$\cdots$ &$728_{-44}^{+48}$ &Equilibrium temperature\\
        \hline
    \label{allpriors}    
    \end{tabular}}
    \begin{tablenotes}
       \item[1]  [1]\ The stellar parameters are adopted from \cite{Gan2023TOI4201}.  
       \item[2]  [2]\ We do not consider heat distribution between the dayside and nightside here and assume albedo $A_B=0$. 
    \end{tablenotes}
\end{table*}

\begin{figure*}[htb]
    \centering
    \includegraphics[width=\linewidth]{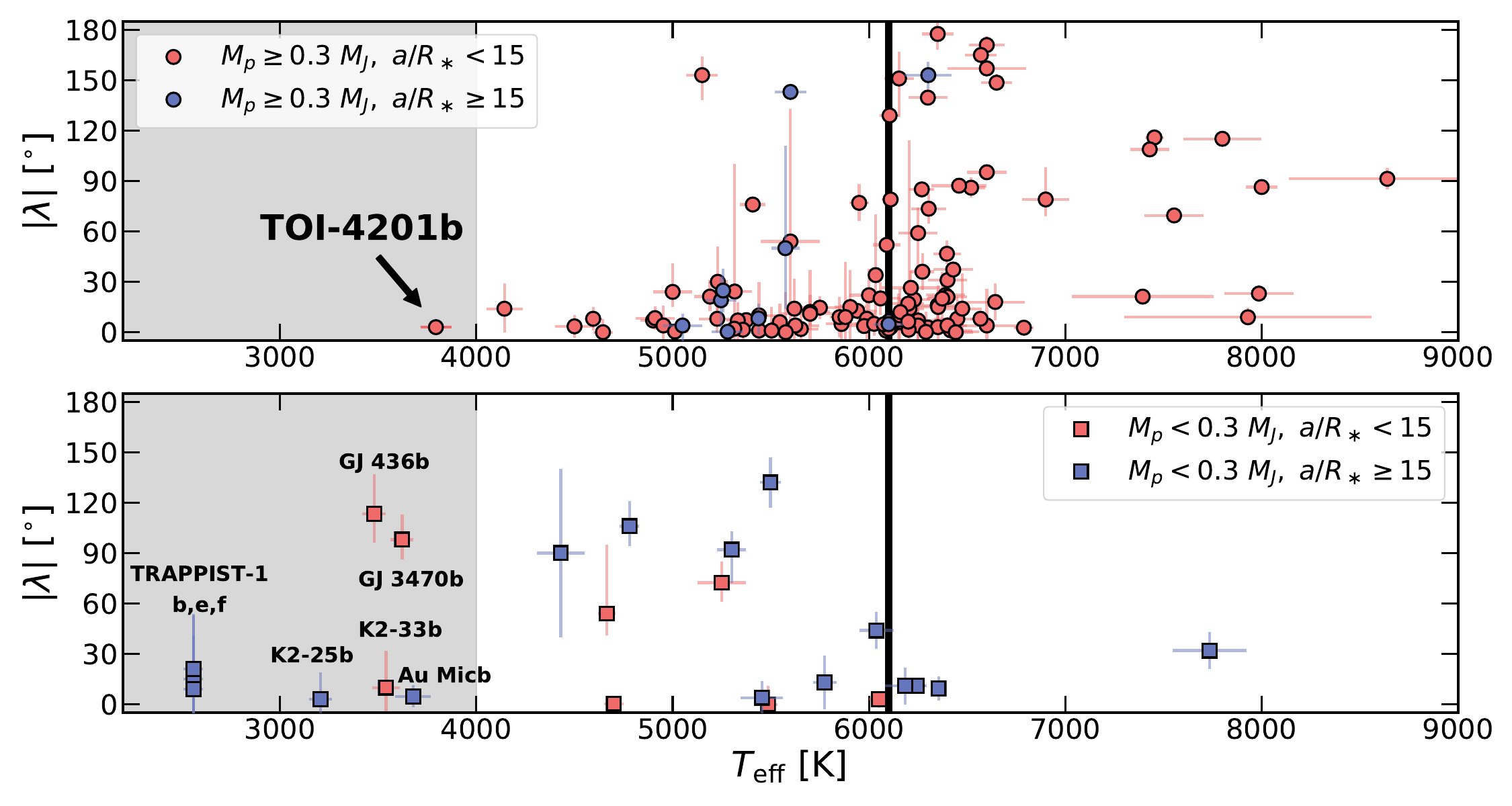}
    \caption{{\it Top panel:} Projected stellar obliquity $\lambda$ of giant-planet systems ($M_p\geq 0.3\ M_J$) vs.\ host star effective temperature. Obliquity results are from RM measurements. The red dots are hot Jupiters (defined as
    having $a/R_\star < 15$)
    and the blue dots are warm Jupiters
    ($a/R > 15$). The position of \tar b is marked with an black arrow. {\it Bottom panel:} Same, but for planets with $M_p<0.3\ M_J$. In both panels, the vertical black line marks the Kraft break. The gray shaded region highlights the
    stars with $T_{\rm eff}\leq 4000$~K. Results were retrieved from the NASA Exoplanet Archive \citep{Akeson2013}.}
    \label{fig:lambda_teff}
\end{figure*}

\section{Discussion}\label{discussion}

\subsection{The Dynamical History of \tar b}



\tar b is one of the most massive hot Jupiters known to exist around an M dwarf. The planet's mass of $2.59^{+0.10}_{-0.11}\ M_J$ is about 5 times heavier than the other known systems ($\sim 0.5\ M_J$). Therefore, the dynamical history of \tar b may be different than those of previously studied systems.


A collision between two planets leading to a merger could account for the high mass, but might tend to misalign the orbit relative to the star \citep{Chatterjee2008}, whereas we have found \tar b to be well aligned. The good alignment might be the result of a more quiescent history, or of tidal obliquity damping following any excitation. Likewise, the planet's low orbital eccentricity argues against high-eccentricity migration unless tidal eccentricity damping has erased the evidence.


As a simple estimate of the tidal realignment timescale $\tau_{\rm CE}$, we used the equation given by \cite{Zahn1977}:
\begin{equation}
    \tau_{\rm CE} = 10\ {\rm Gyr} \left(\frac{M_{p}}{M_{\ast}}\right)^{-2}\left(\frac{a/R_\ast}{40}\right)^{6},
\end{equation}
where $M_{p}/M_\ast$ is the planet-to-star mass ratio and $a/R_\ast$ represents the semi-major axis in units of the stellar radius (see also \citealt{Albrecht2012}). We note that this scaling relation is calibrated using stellar binaries with an assumption that planetary systems have a similar process. With this in mind, we obtain a $\tau_{\rm CE}\approx 950$ Gyr using the updated physical parameters from the joint fit. The realignment timescale is much longer than any astrophysical timescale thus the stellar obliquity is not expected to significantly change after the planet was born. In addition to the dynamical tidal effect timescale above, we also derive the timescale of equilibrium tides using Eq.~2 in \cite{Lai2012}. Adopting a reduced tidal quality factor $Q'$ between $10^{6}$ and $10^{8}$ \citep{Brown2011}, we find that the $\tau_{\rm CE,eq}$ ranges from 2 to 200 Gyr, which is likely beyond the stellar age 0.7-2.0 Gyr estimated by \cite{Gan2023TOI4201} based on the empirical age-rotation relations \citep{Barnes2007,Mamajek2008,Engle2018}. 

Next, we compute the ratio between the rotational angular momentum of the stellar convective layer and the planet's orbital angular momentum to evaluate whether the planet has the capability to realign the host star. The rotational angular momentum of the stellar convective layer is defined as
\begin{equation}
    L_{\rm conv} = \frac{2\pi}{P_{\rm rot}}\times \kappa M_{\rm conv} R_\ast^{2},
\end{equation}
where $\kappa M_{\rm conv} R_{\ast}^{2}$ is the moment of inertia with $\kappa$ depending on the stellar structure and $M_{\rm conv}$ represents the mass of stellar convective layer. The planet orbital angular momentum is calculated through
\begin{equation}
    L_p = \sqrt{G M_\ast M^{2}_{p} a (1-e^{2})}\cos i_p,
\end{equation}
where $M_\ast$ and $M_p$ are the stellar and planet mass, $a$ is the semi-major axis, $e$ and $i_p$ represent the orbital eccentricity and inclination. According to the stellar models built by \cite{Pinsonneault2001}, we find that \tar\ has a convective zone mass $M_{\rm conv}$ of about $0.1\ M_\odot$. Taking $\kappa$ ranging from 0.1 to 0.2 \citep{Baraffe2015}, we obtain an angular momentum ratio $L_{\rm conv}/L_p$ between 0.15 and 0.3, indicating that \tar b may be able to realign the star. To summarize, \tar b has enough angular momentum to realign the spin axis of its host star before being destroyed by tidal decay, but the estimated timescale for realignment is longer than the estimated age of the system, though we note that the timescale estimations both have large uncertainties. Our results suggest that a dynamical quiet history of \tar b is more likely. 


Previous studies on hot Jupiters around FGK stars suggest that tidal obliquity damping is negligible for planets with $a/R_\ast \geq 10$ \citep{Albrecht2022}. By this standard, \tar\ (with $a/R_\ast\approx 13.6$) is not expected to be affected significantly by tidal effects and would have preserved any initial misalignment. The well-aligned orbit of TOI-4201b hints that the limit of $a/R_\ast=10$ on FGK stars may not be the same for M dwarfs. The scaled semi-major axis boundary of strong and weak tidal obliquity damping for M dwarfs is possibly further away ($a/R_\ast>10$). On the other hand, however, \tar\ is less massive than FGK stars, making it easier to be realigned. At this point, we are not able to conclude the role of tidal damping plays in this system. More such measurements on M dwarfs are required to figure out the tidal effect on different planetary systems.







\subsection{\tar b in the Broader Context}\label{four_group_obliquity}

Figure~\ref{fig:lambda_teff} compares the stellar effective temperature sky-projected obliquities of four types of planetary systems, differing in whether the planet is ``hot'' or ``warm'' and whether the planet is a giant or not. The boundary for these categorizations were $a/R_\ast= 15$ and $M_p=0.3\,M_J$, which are chosen somewhat arbitrarily. We note that we did not put any restrictions on the stellar types here. 

For the giant planets, we recover the known results that hot Jupiters
around hot stars show a broader range of obliquities than those around cool stars with effective temperature $T_{\rm eff}< 6,250$ K \citep{Winn2010,Albrecht2012}. Our observation of \tar b has extended the domain of spin-orbit angle studies to include a Jupiter-like planet around an M dwarf. The result is in agreement with the previous findings. The behavior of small planet systems seems more complicated. Below the Kraft break, several low-mass planets are found on nearly polar orbits (e.g., HAT-P-11b, \citealt{Sanchis-Ojeda2011}; WASP-107b, \citealt{Dai2017wasp107,Rubenzahl2021}; HD 3167c, \citealt{Dalal2019,Bourrier2021}; GJ 436b, \citealt{Bourrier2018}; GJ 3470b, \citealt{Stefansson2022}). In fact, short-period small planets around M dwarfs that have deep convective envelopes tend to have a wide range of obliquity, similar to FGK counterparts.

\begin{figure*}
\centering
\includegraphics[width=\linewidth]{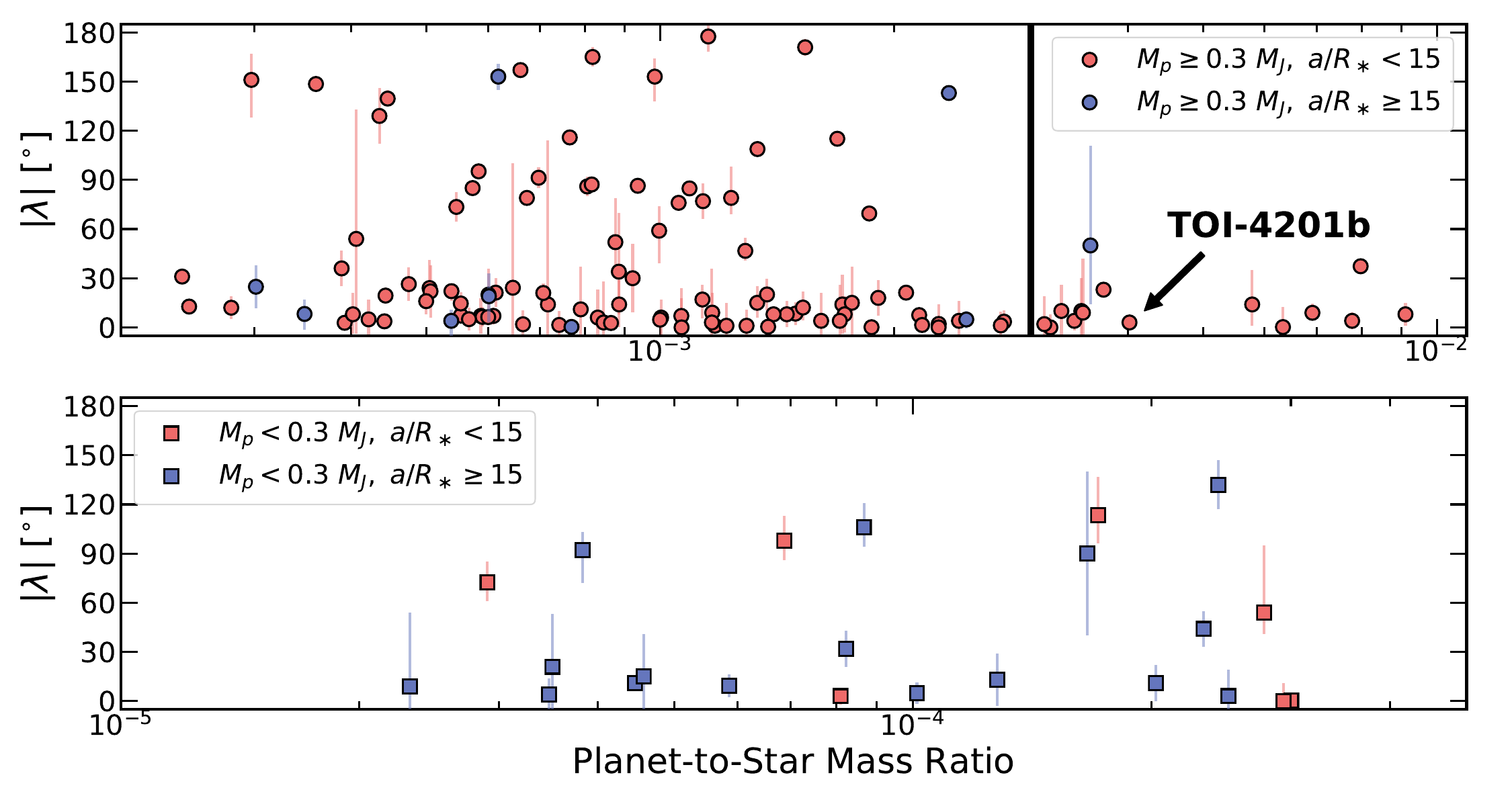}
\caption{Projected stellar obliquity $\lambda$ vs. planet-to-star mass ratio, for four planet groups. Obliquity results are from RM measurements. Different colors represent planets with different scaled semi-major axis $a/R_\ast$. Giant planet systems with mass ratio $M_p/M_\ast$ over $3\times 10^{-3}$ (black vertical line) tend to be aligned while small planets have a broad range of $\lambda$ across mass ratio between $2\times 10^{-5}$ and $3\times 10^{-4}$.  The position of \tar b is marked with an black arrow.}
\label{fig:lambda}
\end{figure*}

In terms of planet-to-star mass ratio, \cite{Hebrard2011} reported that misaligned systems such as retrograde ones usually do not involve the most massive planets. Such correlation still exists around stars with $T_{\rm eff}\leq 7,000$ K but disappears around hotter stars when the sample is doubled \citep{Albrecht2022}. Figure~\ref{fig:lambda} displays the 2D distributions of $\lambda$ and mass ratio $M_p/M_\ast$ of four planet groups. With a mass ratio of about 0.4\%, \tar b is located in the high mass-ratio region where there are relatively few data points.
Based on the figure, it seems that Jupiter-like planets with $M_{p}/M_{\ast}\geq 3\times 10^{-3}$ prefer aligned orbits, although we note that the sample is limited. The stellar obliquities of small planet systems, however, seem to have weak dependence on the mass ratio. Combining the giant and small planet samples, it turns out that the obliquity of planetary systems are mixed below a certain planet-to-star mass ratio. More observations on high-mass-ratio systems are required to draw a firm conclusion.

\section{Conclusions}\label{conclusion}
We reported the MAROON-X spectroscopic observations of a transit of \tar b, and determined that the planet's orbit is well aligned with the host star's spin axis with a sky-projected obliquity of $\lambda=-3.0_{-3.2}^{+3.7}\ ^{\circ}$. Coupled with an estimate of the stellar rotation period, we found the true obliquity to be $\psi=21.3_{-12.8}^{+12.5}\ ^{\circ}$ with an upper limit of $40^{\circ}$ at a 95\% confidence level, suggesting either a dynamically cold formation history or tidal obliquity damping. \tar b is the first hot Jupiter orbiting an M dwarf for which the obliquity has been measured via the RM effect. This observation extends the spin-orbit alignment studies of Jupiter-like planets from FGK to M dwarfs.

We further study \tar b in the context of other systems with measured obliquities. We find that it is consistent with previous findings that hot-Jupiter hosts with $T_{\rm eff}<6{,}250$~K tend to have low obliquities. In addition, \tar b joins the small but growing group of high mass ratio and aligned systems. Planetary systems with mass ratio $\gtrsim 3\times 10^{-3}$ seem to prefer aligned orbits.



\section{Acknowledgments}
We thank Xianyu Wang for useful discussions, and the anonymous referee for the comments that improved the quality of this Letter. This work is partly supported by the National Science Foundation of China (Grant No. 12133005). T. G. acknowledges the Tsinghua Astrophysics High-Performance Computing platform at Tsinghua University for providing computational and data storage resources that have contributed to the research results reported within this Letter. E. P. acknowledges financial support from the Agencia Estatal de Investigaci\'on of the Ministerio de Ciencia e Innovaci\'on MCIN/AEI/10.13039/501100011033 and the ERDF “A way of making Europe” through project PID2021-125627OB-C32, and from the Centre of Excellence “Severo Ochoa” award to the Instituto de Astrofisica de Canarias.

The University of Chicago group acknowledges funding for the MAROON-X project from the David and Lucile Packard Foundation, the Heising-Simons Foundation, the Gordon and Betty Moore Foundation, the Gemini Observatory, and the NSF (award number 2108465).

Based on observations obtained at the international Gemini Observatory, a program of NSF NOIRLab, which is managed by the Association of Universities for Research in Astronomy (AURA) under a cooperative agreement with the U.S. National Science Foundation on behalf of the Gemini Observatory partnership: the U.S. National Science Foundation (United States), National Research Council (Canada), Agencia Nacional de Investigaci\'{o}n y Desarrollo (Chile), Ministerio de Ciencia, Tecnolog\'{i}a e Innovaci\'{o}n (Argentina), Minist\'{e}rio da Ci\^{e}ncia, Tecnologia, Inova\c{c}\~{o}es e Comunica\c{c}\~{o}es (Brazil), and Korea Astronomy and Space Science Institute (Republic of Korea).

This work was enabled by observations made from the Gemini North telescope, located within the Maunakea Science Reserve and adjacent to the summit of Maunakea. We are grateful for the privilege of observing the Universe from a place that is unique in both its astronomical quality and its cultural significance.



%

\vspace{5mm}
\facilities{Gemini-North/MAROON-X, TESS, Keck/HIRES, WIYN/NEID, Magellan/PFS, CFHT/SPIRou, LCOGT, MuSCAT, SPECULOOS}


\software{Allesfitter \citep{Gunther2021}, emcee \citep{Foreman2013}
          }




\appendix

\section{Spectroscopic data obtained for TOI-4201 with MAROON-X}

We present the RVs and activity indicators including CRX, dLW and $S_{\rm H\alpha}$ extracted with \code{SERVAL} \citep{Zechmeister2018} in Table~\ref{RVtable}.

\begin{table}[htb]
    {\renewcommand{\arraystretch}{0.95}
    \caption{Radial velocities and stellar activity indices for \tar\ collected with MAROON-X in this work}
    \begin{tabular}{ccccccccc}
        \hline\hline
        BJD       &RV (m~s$^{-1}$) &$\sigma_{\rm RV}$ (m~s$^{-1}$) &CRX &CRXerr &dLW &dLWerr &$S_{\rm H\alpha}$  &$S_{\rm H\alpha}$err  \\\hline
        MAROON-X Blue & &\\
        2460304.803376 	&81.49 	&4.29 &-92.46 	&62.46 	&21.94 	&7.67 	&0.56 	&0.01 \\
        2460304.815858 	&64.21 	&4.25 &56.85 	&65.47 	&14.53 	&7.61 	&0.55 	&0.01 \\
        2460304.826498 	&63.31 	&4.18 &26.68 	&53.71 	&10.25 	&7.48 	&0.57 	&0.01 \\
        2460304.838082 	&54.83 	&4.15 &-12.55 	&49.57 	&27.16 	&7.42 	&0.56 	&0.01 \\
        2460304.849669 	&47.21 	&4.58 &-105.69 	&80.20 	&34.24 	&8.21 	&0.55 	&0.01 \\
        2460304.861004 	&33.78 	&4.45 &23.22 	&53.74 	&20.13 	&7.99 	&0.55 	&0.01 \\
        2460304.872871 	&62.24 	&4.63 &-143.41 	&83.44 	&-8.68 	&8.35 	&0.54 	&0.01 \\
        2460304.884524 	&44.89 	&4.55 &-85.99 	&70.29 	&27.17 	&8.14 	&0.58 	&0.01 \\
        2460304.895956 	&19.79 	&4.50 &89.52 	&54.11 	&24.90 	&8.05 	&0.56 	&0.01 \\
        2460304.907782 	&-9.03 	&4.61 &133.04 	&83.74 	&30.56 	&8.23 	&0.56 	&0.01 \\
        2460304.919321 	&-36.27 &4.30 &-70.71 	&69.76 	&28.40 	&7.68 	&0.54 	&0.01 \\
        2460304.930921 	&-64.20 &4.58 &156.22 	&77.01 	&22.85 	&8.21 	&0.54 	&0.01 \\
        2460304.942455 	&-56.35 &4.39 &15.18 	&79.36 	&-2.49 	&7.90 	&0.55 	&0.01 \\
        2460304.954451 	&-43.10 &4.59 &-27.47 	&55.60 	&28.76 	&8.23 	&0.55 	&0.01 \\
        2460304.965784 	&-58.58 &4.24 &8.58 	&82.19 	&3.47 	&7.61 	&0.54 	&0.01 \\
        2460304.977139 	&-66.62 &4.96 &-89.77 	&68.64 	&44.29 	&8.87 	&0.56 	&0.01 \\
        2460304.989955 	&-67.93 &5.42 &-12.89 	&86.10 	&50.25 	&9.66 	&0.53 	&0.01 \\
        2460305.000413 	&-80.07 &5.93 &92.30 	&72.43 	&67.84 	&10.53 	&0.56 	&0.01 \\
        2460305.011955 	&-111.20 &6.07 &251.34 	&93.57 	&88.95 	&10.76 	&0.55 	&0.02 \\\hline
         MAROON-X Red & &\\
        2460304.803376 	&71.89 	&4.32 &1.46 	&53.29 	&31.38 	&7.09 	&0.59 	&0.01 \\
        2460304.815858 	&69.97 	&4.29 &19.31 	&55.58 	&42.28 	&7.02 	&0.58 	&0.01 \\
        2460304.826498 	&60.06 	&4.22 &46.96 	&52.14 	&20.67 	&6.91 	&0.59 	&0.01 \\
        2460304.838082 	&52.06 	&4.19 &6.83 	&60.75 	&29.48 	&6.89 	&0.59 	&0.01 \\
        2460304.849669 	&40.59 	&4.59 &-55.61 	&55.97 	&28.89 	&7.56 	&0.56 	&0.01 \\
        2460304.861004 	&30.89 	&4.51 &21.98 	&59.35 	&11.36 	&7.46 	&0.59 	&0.01 \\
        2460304.872871 	&57.89 	&4.68 &61.01 	&60.34 	&7.27 	&7.75 	&0.62 	&0.02 \\
        2460304.884524 	&45.32 	&4.64 &-53.73 	&77.62 	&14.03 	&7.68 	&0.61 	&0.02 \\
        2460304.895956 	&16.50 	&4.51 &-65.55 	&48.81 	&18.67 	&7.44 	&0.58 	&0.01 \\
        2460304.907782 	&-12.43 &4.69 &-22.05 	&74.27 	&40.85 	&7.69 	&0.56 	&0.02 \\
        2460304.919321 	&-40.13 &4.39 &-57.66 	&59.70 	&23.03 	&7.24 	&0.59 	&0.01 \\
        2460304.930921 	&-68.48 &4.64 &9.85 	&48.60  &10.45 	&7.68 	&0.59 	&0.02 \\
        2460304.942455 	&-52.46 &4.48 &-88.21 	&60.17 	&10.12 	&7.44 	&0.58 	&0.01 \\
        2460304.954451 	&-50.40 &4.67 &24.16 	&71.68 	&25.09 	&7.72 	&0.60 	&0.02 \\
        2460304.965784 	&-57.35 &4.34 &16.28 	&79.69 	&29.56 	&7.19 	&0.57 	&0.01 \\
        2460304.977139 	&-76.70 &4.99 &-95.00 	&58.19 	&17.36 	&8.27 	&0.59 	&0.02 \\
        2460304.989955 	&-89.01 &5.43 &-78.90 	&68.27 	&-6.65 	&9.03 	&0.58 	&0.02 \\
        2460305.000413 	&-82.69 &5.91 &147.88 	&76.44 	&9.31 	&9.81 	&0.59 	&0.02 \\
        2460305.011955 	&-85.63 &6.04 &-115.77 	&78.04 	&13.34 	&10.03 	&0.60 	&0.02 \\
         \hline
    \label{RVtable}
    \end{tabular}}
\end{table}




\section{Joint-fit results}
\begin{figure*}[ht]
    \centering
    \includegraphics[width=\linewidth]{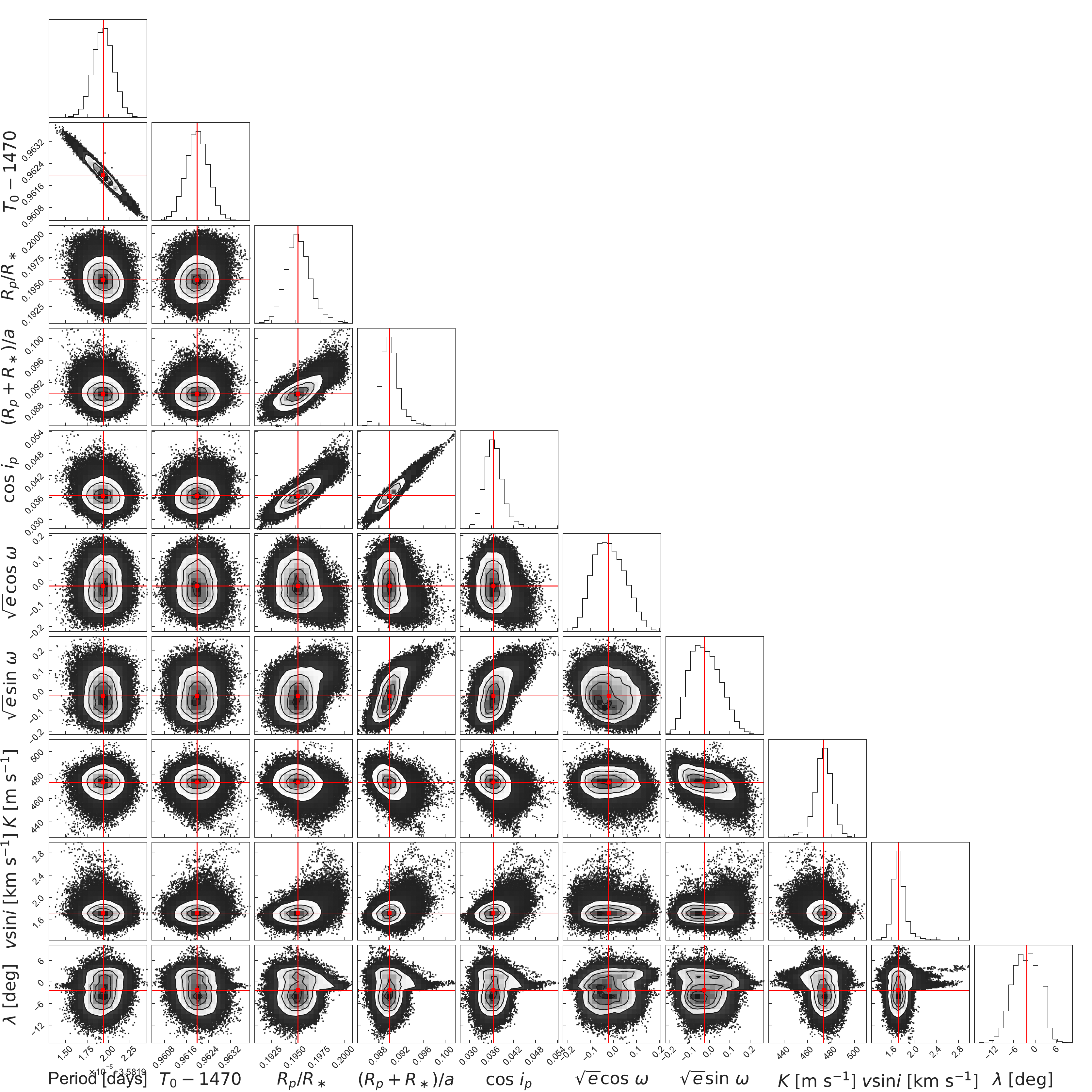}
    \caption{The posterior distributions of key parameters in the joint-fit (see Section~\ref{jointfit}). The red solid lines mark the median values of each distribution.}
    \label{fig:contour}
\end{figure*}

Figure \ref{fig:contour} shows the posterior distributions of key parameters in the joint fit. Table~\ref{allotherpriors} summarizes the prior settings and posteriors of other parameters except for those in Table~\ref{allpriors}. 

\begin{table*}[ht]\scriptsize
    \centering
    {\renewcommand{\arraystretch}{1.05}
    \caption{Priors and best-fits of other parameters in the joint model. $\mathcal{TN}$($\mu$, $\sigma^{2}$, a, b) represents a truncated normal prior ranging from $a$ to $b$.}
    \begin{tabular}{lcc}
        \hline\hline
        Parameter       &Prior &Best-fit   \\\hline
        \it{Dilution factors}\\
        $D_{\rm TESS\ S06}$ &$\mathcal{TN}$ ($0.1$\ ,\ $0.1^{2}$,\ 0\ ,\ 1) &$0.171_{-0.020}^{+0.019}$ \\
        $D_{\rm TESS\ S33}$ &$\mathcal{TN}$ ($0.1$\ ,\ $0.1^{2}$,\ 0\ ,\ 1) &$0.142_{-0.015}^{+0.016}$ \\
        $D_{\rm ground}$ &0 (Fixed) &$\cdots$ \\\hline
        \it{Limb-darkening coefficients}\\
        $q_{\rm 1,TESS\ S06}$ &$\mathcal{U}$ ($0$\ ,\ $1$) &$0.08_{-0.08}^{+0.16}$ \\
        $q_{\rm 2,TESS\ S06}$ &$\mathcal{U}$ ($0$\ ,\ $1$) &$0.67_{-0.33}^{+0.32}$  \\
        $q_{\rm 1,TESS\ S33}$ &$\mathcal{U}$ ($0$\ ,\ $1$) &$0.64_{-0.20}^{+0.22}$  \\
        $q_{\rm 2,TESS\ S33}$ &$\mathcal{U}$ ($0$\ ,\ $1$) &$0.15_{-0.14}^{+0.33}$  \\
        $q_{\rm LCO,SAAO,g}$ &$\mathcal{U}$ ($0$\ ,\ $1$) &$0.76_{-0.32}^{+0.34}$ \\
        $q_{\rm LCO,SAAO,i}$ &$\mathcal{U}$ ($0$\ ,\ $1$) &$0.31_{-0.18}^{+0.23}$  \\
        $q_{\rm LCO,CTIO,g}$ &$\mathcal{U}$ ($0$\ ,\ $1$) &$0.73_{-0.11}^{+0.09}$  \\
        $q_{\rm LCO,CTIO,i}$ &$\mathcal{U}$ ($0$\ ,\ $1$) &$0.29_{-0.11}^{+0.10}$  \\
        $q_{\rm MuSCAT,g}$ &$\mathcal{U}$ ($0$\ ,\ $1$) &$0.55_{-0.07}^{+0.06}$ \\
        $q_{\rm MuSCAT,r}$ &$\mathcal{U}$ ($0$\ ,\ $1$) &$0.62_{-0.04}^{+0.03}$  \\
        $q_{\rm MuSCAT,z}$ &$\mathcal{U}$ ($0$\ ,\ $1$) &$0.33_{-0.05}^{+0.04}$  \\
        $q_{\rm SPECULOOS,z}$ &$\mathcal{U}$ ($0$\ ,\ $1$) &$0.34_{-0.06}^{+0.05}$ \\\hline
        \it{Relative photometric offset}\\
        $M_{\rm TESS\ S06}$ &$\mathcal{U}$ ($-1$\ ,\ $1$) &$0.00013_{-0.00031}^{+0.00030}$\\
        $M_{\rm TESS\ S33}$ &$\mathcal{U}$ ($-1$\ ,\ $1$) &$0.00021_{-0.00022}^{+0.00023}$\\
        $M_{\rm LCO,SAAO,g}$ &$\mathcal{U}$ ($-1$\ ,\ $1$) &$0.00014_{-0.00219}^{+0.00226}$\\
        $M_{\rm LCO,SAAO,i}$ &$\mathcal{U}$ ($-1$\ ,\ $1$) &$0.00080_{-0.00084}^{+0.00085}$\\
        $M_{\rm LCO,CTIO,g}$ &$\mathcal{U}$ ($-1$\ ,\ $1$) &$-0.00039_{-0.00071}^{+0.00070}$\\
        $M_{\rm LCO,CTIO,i}$ &$\mathcal{U}$ ($-1$\ ,\ $1$) &$-0.00024_{-0.00051}^{+0.00054}$\\
        $M_{\rm MuSCAT,g}$ &$\mathcal{U}$ ($-1$\ ,\ $1$) &$0.00492_{-0.00033}^{+0.00032}$\\
        $M_{\rm MuSCAT,r}$ &$\mathcal{U}$ ($-1$\ ,\ $1$) &$0.00297_{-0.00017}^{+0.00016}$\\
        $M_{\rm MuSCAT,z}$ &$\mathcal{U}$ ($-1$\ ,\ $1$) &$0.00364_{-0.00013}^{+0.00014}$\\
        $M_{\rm SPECULOOS,z}$ &$\mathcal{U}$ ($-1$\ ,\ $1$) &$-0.00021_{-0.00020}^{+0.00021}$\\\hline
        \it{Photometric jitter}\\
        $\ln{\sigma_{\rm TESS\ S06}}$  &$\mathcal{U}$ ($-15$\ ,\ $0$)  &$-5.99_{-0.08}^{+0.09}$  \\
        $\ln{\sigma_{\rm TESS\ S33}}$  &$\mathcal{U}$ ($-15$\ ,\ $0$)  &$-5.59_{-0.03}^{+0.04}$  \\
        $\ln{\sigma_{\rm LCO,SAAO,g}}$  &$\mathcal{U}$ ($-15$\ ,\ $0$)  &$-4.73_{-0.16}^{+0.19}$  \\
        $\ln{\sigma_{\rm LCO,SAAO,i}}$  &$\mathcal{U}$ ($-15$\ ,\ $0$)  &$-5.27_{-0.16}^{+0.17}$  \\
        $\ln{\sigma_{\rm LCO,CTIO,g}}$  &$\mathcal{U}$ ($-15$\ ,\ $0$)  &$-5.77_{-0.14}^{+0.16}$  \\
        $\ln{\sigma_{\rm LCO,CTIO,i}}$ &$\mathcal{U}$ ($-15$\ ,\ $0$)  &$-6.16_{-0.15}^{+0.21}$  \\
        $\ln{\sigma_{\rm MuSCAT,g}}$ &$\mathcal{U}$ ($-15$\ ,\ $0$)  &$-5.73_{-0.06}^{+0.07}$  \\
        $\ln{\sigma_{\rm MuSCAT,r}}$ &$\mathcal{U}$ ($-15$\ ,\ $0$)  &$-6.03_{-0.04}^{+0.05}$  \\
        $\ln{\sigma_{\rm MuSCAT,z}}$ &$\mathcal{U}$ ($-15$\ ,\ $0$)  &$-6.29_{-0.04}^{+0.04}$  \\
        $\ln{\sigma_{\rm SPECULOOS,z}}$  &$\mathcal{U}$ ($-15$\ ,\ $0$)  &$-5.51_{-0.03}^{+0.04}$  \\\hline
        \it{Relative RV offset}\\
        $\mu_{\rm HIRES}$ (m~s$^{-1}$) &$\mathcal{U}$ ($-500$\ ,\ $500$) &$3.0_{-13.1}^{+14.3}$\\
        $\mu_{\rm NEID}$ (m~s$^{-1}$) &$\mathcal{U}$ ($-500$\ ,\ $500$) &$-288.1_{-40.5}^{+40.6}$\\
        $\mu_{\rm PFS}$ (m~s$^{-1}$) &$\mathcal{U}$ ($-500$\ ,\ $500$) &$44.1_{-7.7}^{+7.6}$\\
        $\mu_{\rm SPIRou}$ (m~s$^{-1}$) &$\mathcal{U}$ ($-500$\ ,\ $500$) &$-19.9_{-12.0}^{+12.1}$\\
        $\mu_{\rm MAROON-X,Blue}$ (m~s$^{-1}$) &$\mathcal{U}$ ($-500$\ ,\ $500$) &$-1.5_{-4.2}^{+4.4}$\\
        $\mu_{\rm MAROON-X,Red}$ (m~s$^{-1}$) &$\mathcal{U}$ ($-500$\ ,\ $500$) &$-4.8_{-3.5}^{+4.2}$\\\hline
        \it{Spectroscopic jitter}\\
        $\ln{\sigma_{\rm HIRES}}$ (m~s$^{-1}$) &$\mathcal{U}$ ($-15$\ ,\ $0$) &$-2.9_{-0.3}^{+0.4}$\\
        $\ln{\sigma_{\rm NEID}}$ (m~s$^{-1}$) &$\mathcal{U}$ ($-15$\ ,\ $0$) &$-8.3_{-3.5}^{+2.3}$\\
        $\ln{\sigma_{\rm PFS}}$ (m~s$^{-1}$) &$\mathcal{U}$ ($-15$\ ,\ $0$) &$-3.6_{-0.3}^{+0.4}$\\
        $\ln{\sigma_{\rm SPIRou}}$ (m~s$^{-1}$) &$\mathcal{U}$ ($-15$\ ,\ $0$) &$-3.0_{-0.2}^{+0.3}$\\
        $\ln{\sigma_{\rm MAROON-X,Blue}}$ (m~s$^{-1}$) &$\mathcal{U}$ ($-15$\ ,\ $0$) &$-6.8_{-4.8}^{+2.6}$\\
        $\ln{\sigma_{\rm MAROON-X,Red}}$ (m~s$^{-1}$) &$\mathcal{U}$ ($-15$\ ,\ $0$) &$-10.7_{-4.0}^{+2.6}$\\
        \hline
    \label{allotherpriors} 
    \end{tabular}}
\end{table*}

\bibliography{planet}{}
\bibliographystyle{aasjournal}



\end{document}